\documentclass[12pt]{article}
\usepackage[dvips]{graphics}

\begin{document}
\hspace{4.5cm} 

\medskip
\begin{center}

\Large{\bfseries 
The Consequences of the Charge for the Mass of the Elementary Particles.}

\vspace{0.5cm}
\large{E.L. Koschmieder}

\bigskip

\small{Center for Statistical Mechanics\\The University of Texas at 
Austin, Austin TX
78712.  USA\\
e-mail: koschmieder@utexas.edu}

\end{center}

\vspace{0.5cm}

\bigskip

\noindent
\small{We study the consequences which the presence of an 
electric charge e$^\pm$ in the pion $\pi^\pm$ and the 
muon $\mu^\pm$ has for the rest mass of $\pi^\pm$ and 
$\mu^\pm$. The addition of the electric charges 
e$^\pm$ to the massive neutral bodies of these particles 
does not increase the  energy in the rest mass of the pion 
and muon, but rather decreases their energy by the binding 
energy of the electric charge to the 
neutral bodies of the pion and muon. The addition of a charge 
e$^\pm$ to the neutral neutrino lattices of the pion
or muon changes the simple cubic lattices of the neutral particles 
to face-centered cubic lattices of the charged particles, which is
essential for the stability of the particles.}
\normalsize

\smallskip
\noindent
Key words: electron, muon, pion, particle masses, lattice theory.

\section*{Introduction}

\bigskip
\noindent
The electric charge e$^\pm$ is bound to  
the neutral bodies which make up most of the pion and muon 
masses. The mass of the electron or positron is but a
very small part of the mass of $\mu^\pm$ and $\pi^\pm$, 
m($\mu^\pm$) $\cong$ 207\,m(e$^\pm$), m($\pi^\pm$) 
$\cong$ 273\,m(e$^\pm$). 
It has to be explained how the electric charges e$^\pm$ 
affect the mass of the pions and muons.  
We have described in [1] an explanation of the neutral bodies 
of the pions and muons as consisting of simple cubic lattices made 
up by muon neutrinos, 
anti-muon neutrinos, electron neutrinos and anti-electron neutrinos  
and their oscillations. In Section\,11 in [1] we have also explained the 
rest mass of the electron or positron as consisting of an electron neutrino 
or anti-electron neutrino lattice and of electric oscillations. The accurately
known ratios m($\mu^\pm$)/m(e$^\pm$) and m($\pi^\pm$)/m(e$^\pm$)
seem to be the most sensitive gauge for the influence of the charge 
on the mass of the particles.     
Any valid model of the elementary particles must, in the end, be able  
to determine the ratios of the particle masses to the electron mass. 
In particular the one hundred years old problem of why is the mass 
of the proton 1836 times larger than the mass of the electron has to be 
solved.

  The ratios m($\mu^\pm$)/m(e$^\pm$) and m($\pi^\pm$)/m(e$^\pm$)   
were first discussed by Nambu [2] who noted that m($\mu^\pm$)  
is approximately 3/2$\alpha_f\,\cdot\,$m(e$^\pm$) and m($\pi^\pm$) is 
approximately 2/$\alpha_f\,\cdot\,$m(e$^\pm)$, where 
$\alpha_f$ = e$^2$/$\hbar$c $\cong$ 1/137 is the fine structure constant. 
More accurate ratios 
m($\mu^\pm$)/m(e$^\pm$) and m($\pi^\pm$)/m(e$^\pm$) were later suggested 
by Barut [3]. We have shown in Section\,12 of [1] that fairly accurate values 
of the ratios m($\mu^\pm$)/m(e$^\pm$) and m($\pi^\pm$)/m(e$^\pm$) 
can be obtained from our explanation of the masses of the muons, pions 
and electron. Even when the consequences of the charge of 
$\mu^\pm$ and $\pi^\pm$ were neglected we arrived at values for the mass 
ratios which differed from the measured m($\mu^\pm$)/m(e$^\pm$) by only 
0.38\% and from the measured m($\pi^\pm$)/m(e$^\pm$) by 1.08\%. We will 
now improve our calculations of the ratios m($\mu^\pm$)/m(e$^\pm$) and 
m($\pi^\pm$)/m(e$^\pm$) by taking also into account the consequences of 
the charge on the mass of the particles.

\bigskip
\section{The charge in the $\pi^\pm$\,mesons}

   When a $\pi^\pm$\,meson is formed N/4 electron neutrinos $\nu_e$ 
or anti-electron neutrinos $\bar{\nu}_e$, as well as N/4 electric oscillations 
come, according to Eq.(85) in [1], with the electron or positron into 
the $\pi^\pm$ lattice. N/4 electron neutrinos or anti-electron 
neutrinos make up 1/2 of the mass of the electron. 
The N/2 electric oscillations or N/4 charge elements Q$_k$ make up the other 
half of the mass of the electron, as the  
explanation of the electron, we have put forward in [1], demands.
 The number N of all neutrinos in the neutrino lattice of the 
$\pi^\pm$\,mesons without their charge is 
\begin{equation} \mathrm{N} = 2.854\cdot10^{\,9} \,, \end{equation}
Eq.(17) in [1]. N is an absolute number, neutrinos and antineutrinos 
are counted equally.
 
The N/4 electron neutrinos $\nu_e$ originating from the electron
cannot be part of the cubic $\nu_\mu,\bar{\nu}_\mu,\nu_e,\bar{\nu}_e$ 
lattice of the $\pi^\pm$ mesons, because the corners of the cubic cells 
are already filled with neutrinos. The additional electron- or 
anti-electron neutrinos must sit in the centers of the cell-sides.
Likewise, the N/4 charge elements Q$_k$
can only sit in the centers of the cell-sides, because the corners of 
the cubic cells of the neutrino lattice are filled with neutrinos. 
The cells would then be \emph{face-centered cubic}, Fig.\,1. It is
fundamental for the stability of the $\pi^\pm$\,mesons that their
lattice is face-centered. The lifetime of the $\pi^\pm$\,mesons is
2.6$\cdot$10$^{-8}$ sec, whereas the lifetime of the $\pi^0$\,meson,
which does not carry a charge, but has also a simple cubic lattice, 
is 8.4$\cdot$10$^{-17}$ sec. 
Lattice theory finds that simple cubic lattices are not stable, whereas 
face-centered lattices are stable, which is reflected in the lifetime of 
the particles.

\hspace{2.5cm}

\begin{center}

\includegraphics{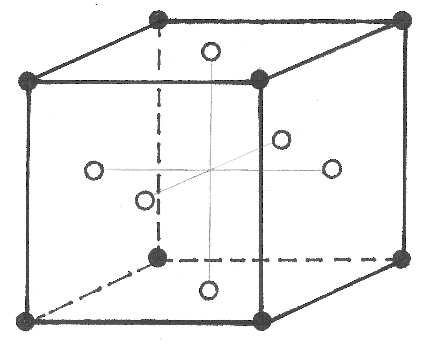}

\begin{quote}
Fig.\,1: A face-centered cubic cell. The cells of a face-centered lattice
can have, but do not have to have, a particle in the center of each 
side of the cells. The thin lines connecting the points at the centers
of the cell-sides are only a visual aid.  
\end{quote}
\end{center}

\bigskip 

   The neutral cubic neutrino lattices of $\mu^\pm$ and $\pi^\pm$  
seem to exist only in conjunction with an electric charge e$^\pm$.
As noted on
p.\,18 of [1], a simple cubic lattice held together by a
central force between the particles of the lattice is not stable
according to Born [4]. A cubic lattice consisting only of neutrinos
falls into this category. The charge e$^\pm$ added to the neutral neutrino 
lattices of $\mu^\pm$ and $\pi^\pm$ makes their neutrino lattices
``stable", because the neutrino lattices are then no longer simple 
cubic, but face-centered cubic. \emph{Face-centered cubic} lattices 
are stable according to Born [4].

\bigskip

 The N/4 charge elements Q$_k$ of a free charge 
e$^\pm$ brought by e$^\pm$ into the neutrino lattice of the pion must
remain intact, because the sum of the charge elements is equal 
to the electric charge e$^\pm$, which must be conserved. Where then 
do the charge elements of e$^\pm$ sit in the lattice of the 
$\pi^\pm$\,mesons\,? According to our explanation of the electron in [1] the charge elements of e$^-$, and the
electron neutrinos $\nu_e$ in e$^-$, form a simple cubic lattice, 
Fig.\,8 in [1]. If this 
structure is maintained when a charge e$^\pm$ attaches to the neutral cubic 
lattice of the $\pi^\pm$\,meson, the charge elements should sit in the 
center of cell-sides which are perpendicular to the sides of the cells on 
which the additional electron- or anti-electron neutrinos sit. The charged 
cubic cells of the $\pi^\pm$ lattice would then be face-centered, as on Fig.\,1.    

\bigskip

The N/4 neutrinos and the N/4 charge elements Q$_k$ of the charge of 
$\pi^\pm$ cannot oscillate with the frequency of the neutrinos in the 
cell corners. The oscillation energy of the $\pi^\pm$ lattice would then 
be proportional to 6\,$\cdot$\,N/4, which is not 
posible, because the oscillation energy of the $\pi^\pm$\,mesons  
would then exceed by 50\% the oscillation energy of the neutral neutrino 
lattice, which is proportional to 4\,$\cdot$\,N/4.   
The additional neutrinos and the charge elements in the centers of the 
cell-sides oscillate with the same frequencies they have in e$^\pm$,
namely with a frequency proportional to $\alpha_f\nu_0$, as required   
by the oscillation frequencies in the electron or positron,  
given in Eq.(79) in [1]
\begin{equation} \mathrm{E}_\nu(\mathrm{e}^\pm) =  
\frac{\mathrm{Nh}\nu_0\cdot\alpha_f}
{(\mathrm{e^{h\nu/kT}}\,
\mathrm{-}\,1)}\cdot\frac{1}{2\pi}\int\limits_{-\pi}^{\pi}\,\phi\,d\phi =
 \mathrm{N}\cdot\frac{\mathrm{e}^2}{a}\cdot\frac{1}{\mathrm{f(T)}}
\cdot\frac{1}{2\pi}\int\limits_{-\pi}^{\pi}\phi\,d\phi = 
1/2\cdot\mathrm{m(e^\pm})\mathrm{c}^2\,.\end{equation}
In Eq.(2) h is Planck's constant, the frequency 
$\nu_0$ is given by $\nu_0$ = c/2$\pi$\emph{a}, where c is the velocity 
of light, \emph{a} is the lattice constant, and 
$\phi$ = 2$\pi$\emph{a}/$\lambda$.
The term e$^2$/\emph{a} in the second part of Eq.(2) comes from the 
relation h$\nu_0$$\alpha_f$ = e$^2$/\emph{a}. There are N/2 electric 
oscillations in e$^\pm$ or N/4 charge elements. Because the fine structure 
constant $\alpha_f$ is $\cong$ 1/137, the oscillation energy in the 
electron adds little to the oscillation energy of the pions, E$_\nu(\pi^\pm$).  

   As we have shown in [1] the ratio m($\pi^\pm$)/m(e$^\pm$) is, 
not considering the consequences of the 
charge on the mass of $\pi^\pm$, given by Eq.(103) therein
\begin{equation} \frac{\mathrm{m}(\pi^\pm)}{\mathrm{m(e^\pm)}}(theor)
= 2\,[\frac{\mathrm{m}(\nu_\mu)}{\mathrm{m}(\nu_e)} +1] 
 \cong \frac{2}{\alpha_f} + 2 = 276.072 = 1.0107\,\frac{\mathrm{m}(\pi^\pm)}
{\mathrm{m(e}^\pm)}(exp)\,, \end{equation} with 
\begin{equation}\mathrm{m}(\nu_e) = \alpha_f\cdot\mathrm{m}(\nu_\mu)\,,
\end{equation} from Eq.(72) in [1].
The empirical formula of Nambu [2] for the
ratio m($\pi^\pm)$/m(e$^\pm$) is 
\begin{equation}\frac{\mathrm{m}(\pi^\pm)}{\mathrm{m(e}^\pm)}(\emph{emp}) = 
\frac{2}{\alpha_f} - 1 = 273.072 = 0.99978\,\frac{\mathrm{m}(\pi^\pm)}
{\mathrm{m(e}^\pm)}(\emph{exp})\,.
\end{equation}
The experimental ratio for m($\pi^\pm$)/m(e$^\pm$) is  
\begin{equation} 
\mathrm{m}(\pi^\pm)/\mathrm{m(e}^\pm)(exp) \cong 273.132\,, \end{equation}
with m($\pi^\pm$)c$^2$ = 139.570\,MeV and m(e$^\pm$)c$^2$ = 
0.5109989\,MeV.

\bigskip 

   The charge elements Q$_k$ brought by e$^\pm$
into the neutral neutrino lattice of the pions add to the 
oscillation energy in the pion lattice. The oscillation energy in the sum 
of the charge elements is equal to

\medskip

\hspace{2cm} $\Sigma_k$Q$_k$ = 1/2\,$\cdot$\,m(e$^\pm$)c$^2$ = 
N/4\,$\cdot$\,m($\nu_e$)c$^2$,

\medskip
\noindent
from Eq.(84) in [1]. On the other hand, there may be a binding energy 
when an electron or positron is brought into the neutrino lattice 
of the neutral pions. The oscillation 
energy of the neutrino lattice of the $\pi^\pm$\,mesons is, Eq.(34) of [1], 
\begin{equation} \mathrm{E}_\nu(\pi^\pm_0) =
\mathrm{N}/2\cdot[\mathrm{m}(\nu_\mu) + \mathrm{m}(\nu_e)]\mathrm{c}^2\,,
\end{equation}
the subscript 0 in $\pi^\pm_0$ means that this deals with the pion without its charge, 
in other words with the neutrino lattice of $\pi^\pm$.
The oscillation energy in the electron or positron is
\begin{equation} \mathrm{E}_\nu(\mathrm{e}^\pm) = 
1/2\cdot\mathrm{m(e}^\pm)\mathrm{c}^2
= \mathrm{N}/4\cdot\mathrm{m}(\nu_e)\mathrm{c}^2
= \mathrm{N}/4\cdot\mathrm{m}(\bar{\nu}_e)\mathrm{c}^2\,, 
\end{equation}
from  Eq.(84) in [1]. Using
\begin{equation}\mathrm{m}(\bar{\nu}_\mu) = \mathrm{m}(\nu_\mu)
\quad \mathrm{and} \quad \mathrm{m}(\bar{\nu}_e) = \mathrm{m}(\nu_e)\,,
\end{equation}
from Eqs.(68,71) in [1], we arrive, from the sum of Eq.(7) and Eq.(8), 
at the oscillation energy in the \emph{charged} pions  
\begin{equation} \mathrm{E}_\nu(\pi^\pm) = 
\mathrm{N}/2\cdot[\mathrm{m}(\nu_\mu) + \mathrm{m}(\nu_e)]\mathrm{c}^2
+ \mathrm{N}/4\cdot \mathrm{m}(\nu_e)\mathrm{c}^2\,.
\end{equation}

\bigskip
 
   The sum of the masses of the 
$\nu_\mu,\,\bar{\nu}_\mu,\,\nu_e,\,\bar{\nu}_e$ 
neutrinos in the lattice of the neutral pions is 
$\Sigma_i$\,m(neutrinos)($\pi^\pm_0$)
= N/2\,$\cdot$\,[m($\nu_\mu$) + m($\nu_e$)], from Eq.(33) in [1].
Adding to that the N/4 \,$\nu_e$ or $\bar{\nu}_e$ neutrinos which come with
the charge e$^\pm$ into the lattice, we find    
the energy in the sum of the neutrino masses in the \emph{charged} pions  
\begin{equation} \Sigma_i\,\mathrm{m(neutrinos)}(\pi^\pm)\mathrm{c}^2 =
[\mathrm{N}/2\cdot\mathrm{m}(\nu_\mu) 
+ \mathrm{N}/2\cdot\mathrm{m}(\nu_e)]\mathrm{c}^2 
+ \mathrm{N}/4\cdot\mathrm{m}(\nu_e)\mathrm{c}^2 \,. \end{equation}
 
   Adding to the oscillation energy in the neutral neutrino lattice of the 
pion E$_\nu(\pi^\pm_0$) (Eq.7) the energy in the sum of the  
masses of the neutrinos in the neutral pion lattice, 
N/2\,$\cdot$\,[m($\nu_\mu$) + m$(\nu_e$)]c$^2$, 
and the energy in the mass of e$^\pm$, which is m(e$^\pm$)c$^2$ =
N/2\,$\cdot$\,m($\nu_e$)c$^2$, Eq.(86) in [1], we find for  
the total energy m($\pi^\pm$)c$^2$ of the \emph{charged} pions 
\begin{eqnarray}\mathrm{m(\pi^\pm)c}^2(theor) &=&
\mathrm{E_\nu(\pi^\pm_0)} 
+ \Sigma_i\,\mathrm{m(neutrinos)(\pi^\pm_0)c}^2 
+ \mathrm{m(e^\pm)}\mathrm{c}^2\nonumber\\&=&  
\mathrm{E_\nu(\pi^\pm_0)} +
 \mathrm{N}/2\cdot[\mathrm{m}(\nu_\mu) + \mathrm{m}(\nu_e)]\mathrm{c}^2
+ \mathrm{N/2}\cdot\mathrm{m}(\nu_e)\mathrm{c}^2 \nonumber\\ &=&
 \mathrm{N}\mathrm{m}(\nu_\mu)\mathrm{c}^2 + \mathrm{3N/2\cdot m}
(\nu_e)\mathrm{c}^2\,,       
\end{eqnarray}
making use of
E$_\nu(\pi^\pm_0)$  =  $\Sigma_i\,\mathrm{m(neutrinos)(\pi^\pm_0)c}^2$ 
in the neutrino lattice of the pion, and of m(e$^\pm$)c$^2$ =
N/2\,$\cdot$\,m($\nu_e$)c$^2$.

\bigskip  

   After division by m(e$^\pm$)c$^2$ 
it follows from Eq.(12) that the ratio of the mass of the charged
pions to the mass of the electron or positron is, using also Eq.(4),
\begin{equation}\frac{\mathrm{m}(\pi^\pm)}{\mathrm{m(e}^\pm)}(theor)
= 2\frac{\mathrm{m}(\nu_\mu)}{\mathrm{m}(\nu_e)} + 3
= \frac{2}{\alpha_f} + 3 = 277.072 
=1.0144\,\frac{\mathrm{m}(\pi^\pm)}{\mathrm{m(e}^\pm)}(exp)\,. 
 \end{equation}
Nambu's empirical ratio m($\pi^\pm$)/m(e$^\pm$) is 2/$\alpha_f$\,$-$\,1 
= 273.072, Eq.(5); the experimental mass ratio is 273.132, Eq.(6). The 
term 2/$\alpha_f$ in Eq.(13) comes from the two neutrinos, a muon neutrino 
and an antimuon neutrino, which are in opposite corners of the cell-sides 
of the cubic neutrino lattice of $\pi^\pm$. The two 
$\nu_\mu$,\,$\bar{\nu}_\mu$ neutrinos make up nearly the entire mass in a 
cell-side. In all cell-sides are then N/2\,$\cdot$\,$\nu_\mu$($\bar{\nu}_\mu$) 
neutrinos. Since the oscillation energy 
is equal to the energy in the sum of the neutrino masses, the energy in 
the lattice is proportional to N\,$\cdot$\,$\nu_\mu$($\bar{\nu}_\mu$). This
divided by m(e$^\pm$) = N/2\,$\cdot$\,m($\nu_e$)($\bar{\nu}_e$) gives,
with Eq.(4), the 2/$\alpha_f$ term.

From the difference between the experimental mass of the pions and 
the theoretical mass of the charged pions follows the binding energy of 
the electric charge. \emph{The binding energy of the electron or positron} to the neutrino lattice of the pions follows from the difference between 
Eq.(13) and Eq.(6) and is  
\begin{equation} \Delta\mathrm{E_{\,b}(\pi^\pm_0,e^\pm}) =
3.94\,\mathrm{m(e^\pm)}\mathrm{c^2} 
\cong 4\,\mathrm{m(e^\pm)c}^2 \,. \end{equation}

\bigskip

   The binding energy can come only from the oscillation energy of the 
neutrinos in the corners of the cell-sides in $\pi^\pm$, because of 
conservation of neutrino numbers, and because the oscillation energy 
of the charge elements brought by e$^\pm$ into $\pi^\pm$ must be 
conserved, since the sum of the charge elements represents the charge of 
the electron. The introduction of a neutrino and a charge element into the 
center of the sides of the cells of the cubic neutrino lattice causes a 
reduction of the oscillation frequencies of the neutrinos in the corners 
of the cell-sides.

\bigskip      
 
   We now subtract the binding energy in Eq.(14), divided by the energy 
in the rest mass of the electron, that means $\Delta$E$\,_b$
($\pi^\pm_0$,\,e$^\pm$)/m(e$^\pm$)c$^2$  = 4,
from Eq.(13) and find that the ratio of the mass of the charged pion to the
mass of the electron is, after the binding energy is taken into account, 

\begin{equation} \mathrm{m}(\pi^\pm)/\mathrm{m(e}^\pm)(theor) = 
2/\alpha_f - 1\,, \end{equation}

\noindent
the same as given by Nambu's formula (Eq.5). It also follows that
\begin{eqnarray}\mathrm{m}(\pi^\pm)\mathrm{c}^2(theor) = 
(2/\alpha_f - 1)\cdot\mathrm{m(e^\pm)c^2} = 
139.5395\,\mathrm{MeV}\nonumber\\ 
= 0.99978\,\mathrm{m}(\pi^\pm)\,\mathrm{c}^2(exp)\,,\end{eqnarray}
or that we have explained the mass of the charged pions with an
accuracy of 2.2\,$\cdot$\,10$^{-4}$.

In simple terms, Eq.(103) in [1] gives the ratio of the mass of the 
neutrino lattice of the neutral pion to the mass of e$^\pm$
\begin{equation} 
\mathrm{m}(\pi^\pm_0)/\mathrm{m(e}^\pm)(theor) \cong 2/\alpha_f +2\,.
 \end{equation}
Adding to the right hand side of this equation + 1 for the mass of
the electron or positron added to the neutrino lattice of the neutral pions, 
and deducting 4 for the binding energy of e$^\pm$ to the neutrino
lattice of the neutral pions, gives Nambu's empirical formula 
2/$\alpha_f$ $-$ 1, Eq.(5), for the ratio of the mass of the charged pions to the
mass of the electron or positron, as it must be.

\bigskip

   To summarize: We have found that the electron neutrinos or 
anti-electron neutrinos brought by e$^\pm$ into the neutral neutrino 
lattice of the pions must sit in the center of the sides, e.g.\,\,the front 
and rear sides, of the  simple cubic cells in the neutral neutrino lattice 
of the pions, and that the charge 
elements sit in the centers of cell-sides which are perpendicular to the 
sides with the $\nu_e$ or $\bar{\nu}_e$ neutrinos coming from
e$^\pm$. That means that the cells in the $\pi^\pm$ lattice 
are face-centered cubic. And we have 
found that bringing e$^\pm$  into the neutral neutrino lattice of the  
pions releases a binding energy which is $\approx$
4\,m(e$^\pm$)c$^2$, and that then the mass m($\pi^\pm$) 
has been explained with an accuracy of 2.2\,$\cdot$\,10$^{-4}$.

\bigskip
\noindent
\section{The charge in the $\mu^\pm$ muons}

\bigskip
   
   In [1] we have shown that the masses of the muons and pions 
can be explained, within 1\% accuracy, by  
the sum of the masses of electron neutrinos and muon neutrinos  
and their antiparticles in simple cubic lattices, and by the mass 
in the energy of the oscillations of the neutrinos.  
Similarly the electron mass can be explained by the 
sum of the masses of electron neutrinos in a cubic lattice 
plus the mass in the sum of the energy 
of electric oscillations. The 
oscillation energy of the neutrinos in the neutral cubic lattice  
of the muons and pions is given by Eqs.(63) and (78) in [1]
\begin{equation}\mathrm{E}_\nu(\mu^\pm_0) = 
\mathrm{E}_\nu(\pi^\pm_0) = 
\frac{\mathrm{Nh}\nu_0}{(e^{h\nu/kT}\,\mathrm{-}\,1)}
\cdot \frac{1}{2\pi}\int\limits_{-\pi}^{\pi}f(\phi)d\phi\,.
\end{equation}
Although the oscillations of a
neutrino lattice are neutral, we must retain the $\pm$ superscript
of $\mu$ and $\pi$ in Eq.(18) because the neutrino lattices of 
e.g.\,\,$\mu^+$ and $\mu^-$  
are not the same, but they have the same oscillation energy. 

   In $\mu^\pm$ an electric charge e$^\pm$ is bound  
to the massive neutral body of the muons, which is about 
205.768\,m(e$^\pm$). The body of the muons contains the
remains of the body of the pions from which the muons
emerge in the decay $\pi^\pm \rightarrow \mu^\pm + 
\nu_\mu(\bar{\nu}_\mu)$. In a good approximation the body of the
muons is 3/4 of the body of the pions, m($\mu^\pm$) = 
1.00937\,$\cdot$\,3/4\,$\cdot$\,m($\pi^\pm$). Either all $\nu_\mu$
or all $\bar{\nu}_\mu$ neutrinos leave in the $\pi^\pm$ decay, the neutral 
body of the muons consists therefore of 3/4\,$\cdot$\,N neutrinos. 
The composition of $\mu^\pm$ is given by Eq.(66) in [1]. This 
equation does not consider the consequences of the charge on the  
mass of $\mu^\pm$, it reads
\medskip
\noindent
\begin{equation} \mathrm{m}(\mu^\pm_0) = 
\mathrm{E}_\nu(\mu^\pm_0)/\mathrm{c}^2 
 + \mathrm{N}/4\cdot\mathrm{m}(\nu_\mu) + 
\mathrm{N}/4\cdot\mathrm{m}(\nu_e) + 
\mathrm{N/4\cdot m}(\bar{\nu}_e)\,, \end{equation}

\medskip
\noindent
or correspondingly with the antiparticles.  m($\nu_\mu$) = 
49.91\,milli-eV/c$^2$ is the
mass of the muon neutrino, and m($\nu_e$) = 0.365\,milli-eV/c$^2$  
is the mass of the electron neutrino, Eqs.(67,70) in [1], both can be
exchanged by their antiparticles. Actually, conservation of 
neutrino numbers requires that N/8 muon neutrinos and N/8 anti-muon 
neutrinos leave the $\pi^\pm$ lattice in the decay of $\pi^\pm$.
Eq.(19) means that the rest mass of the muon m$(\mu^\pm$)\emph(exp)  
= 105.658\,MeV/c$^2$ is equal to the oscillation energy 
E$_\nu(\mu^\pm)$/c$^2$ of the neutrinos in the lattice of the muons, 
plus the sum of the masses of the neutrinos in this lattice.  

   The oscillation energy in the neutral muon lattice is the same as 
the oscillation energy in the neutral pion lattice, as follows from the measured
masses of the muons and pions. According to Eq.(63) in [1] we have
\begin{equation} \mathrm{E}_\nu(\pi^\pm_0) = \mathrm{E}_\nu(\mu^\pm_0)\,.
 \end{equation}
The formula for the oscillation energy of a neutral neutrino lattice, 
Eq.(18), 
applies to the neutrino lattice of the muons only if N neutrinos are in 
$\mu^\pm$, not the 3/4\,$\cdot$\,N neutrinos one expects 
after either N/4 muon neutrinos or N/4 anti-muon neutrinos have 
been removed from the lattice of the pions in the $\pi^\pm$ decay 
$\pi^\pm \rightarrow \mu^\pm + \nu_\mu(\bar{\nu}_\mu)$.   
Applied to the \emph{neutral} lattices of $\mu^\pm$ and $\pi^\pm$ the 
difference in the 
number of the neutrinos is expressed by Eq.(64) in [1] which reads
\begin{equation} \mathrm{m}(\pi^\pm_0) - \mathrm{m}(\mu^\pm_0) 
= \mathrm{N}/4\cdot\mathrm{m}(\nu_\mu) = 
\mathrm{N}/4\cdot\mathrm{m}(\bar{\nu}_\mu) \,. \end{equation}

   According to our explanation of the electron in Section\,11 of [1]  
the charge e$^-$ carries N/4 electron neutrinos and the 
positron carries carries N/4 anti-electron neutrinos.  
When a charge e$^\pm$ is added to the neutral lattice of the muon with 
3/4\,$\cdot$\,N neutrinos, the N/4 electron- or anti-electron neutrinos 
of e$^\pm$ fill the lattice points made vacant by the emission of the muon- 
or antimuon neutrinos in the $\pi^\pm$ decay. With the N/4 neutrinos coming
from e$^\pm$ there are then N neutrinos in the lattice of the charged muon,
that means that the lattice is cubic, and Eq.(18) applies.  
It is likely that the N/4 charge elements brought also by e$^\pm$ 
into the neutral, cubic neutrino lattice of the muon occupy the centers of 
the cell-sides. The charge elements sitting in the centers of the cell-sides 
of the $\pi^\pm$ lattice do apparently not change their position during the
conversion from the $\pi^\pm$ lattice to the $\mu^\pm$ lattice in 
$\pi^\pm$ $\rightarrow$ $\mu^\pm$ +$\nu_\mu(\bar{\nu}_\mu)$. The lattice 
of $\mu^\pm$ would then be face-centered cubic. We must now add 
the energy in the sum of the charge elements, which is according to 
Eq.(84) in [1] equal to 1/2\,$\cdot$\,m(e$^\pm$)c$^2$ or equal to 
N/4\,$\cdot$\,m($\nu_e$)c$^2$, and the energy in the 
N/4\,$\cdot$\,m($\nu_e$)c$^2$ neutrinos in the electron, to the energy in 
the muon, Eq.(19), in order to obtain the energy in the charged muon.

\bigskip
  
   The oscillation energy of the N neutrinos in the lattice of the   
pions is given by Eq.(7). The oscillation energy E$_\nu(\pi^\pm_0)$ is
equal to the energy in the sum of the neutrino masses in $\pi^\pm$,
that means equal to N/2\,$\cdot$\,[m($\nu_\mu)$ + m($\nu_e$)]c$^2$, Eq.(7).
The oscillation energy E$_\nu$($\mu^\pm_0)$ of the neutrino lattice of  
the neutral muons is the same as in $\pi^\pm$ according to Eq.(20). To 
E$_\nu$($\mu^\pm_0)$ we now add the oscillation energy in the electron 
or positron, which is given by Eq.(8)
\begin{equation} \mathrm{E}_\nu(\mathrm{e}^\pm) = 
1/2\cdot\mathrm{m(e}^\pm)\mathrm{c}^2
= \mathrm{N}/4\cdot\mathrm{m}(\nu_e)\mathrm{c}^2
= \mathrm{N}/4\cdot\mathrm{m}(\bar{\nu}_e)\mathrm{c}^2\,, 
\end{equation}
The oscillation energy in a free 
electron or positron must be preserved when they are transferred to
the neutral neutrino lattice of $\mu^\pm$, because the oscillation 
energy in the electron or positron represents the electric charge. 
The oscillation energy in the charged muons is then   
\begin{equation} \mathrm{E}_\nu(\mu^\pm) = 
\mathrm{N}/2\cdot[\mathrm{m}(\nu_\mu) + \mathrm{m}(\nu_e)]
\mathrm{c}^2
+ \mathrm{N}/4\cdot\mathrm{m}(\nu_e)\mathrm{c}^2\,,
 \end{equation}
or we have 
\begin{equation} \mathrm{E}_\nu(\mu^\pm) =  
\mathrm{N}/2\cdot\mathrm{m}(\nu_\mu)\mathrm{c}^2 + 
3/4\cdot\mathrm{N}\,\mathrm{m}(\nu_e)\mathrm{c}^2\,.
 \end{equation}

\bigskip 

   We must now add to the oscillation energy of the charged 
muons, Eq.(24), the energy in the sum of the masses of the neutrinos in 
the charged muon lattice. The sum of the neutrino masses 
in the charged lattice of the muons is equal to the sum of the neutrino 
masses in Eq.(19) plus N/4\,$\cdot$\,m($\nu_e$) from the charge.
In other words, in $\mu^\pm$ are N/4 muon neutrinos $\nu_\mu$ plus N/4 
electron neutrinos $\nu_e$ and N/4 anti-electron neutrinos $\bar{\nu}_e$, 
which come from the $\pi^\pm$ lattice, and N/4 electron- or anti-electron 
neutrinos which come from e$^\pm$. In total N neutrinos are in the charged
cubic $\mu^\pm$ lattice. That means that  
\begin{equation}\Sigma_i\,\mathrm{m(neutrinos)(\mu^\pm)c^2} = 
\mathrm{N/4}\cdot\mathrm{m}(\nu_\mu)\mathrm{c}^2 + 
3/4\cdot\mathrm{N}\,\mathrm{m}(\nu_e)\mathrm{c}^2\,,
\end{equation}
with
\begin{equation}\mathrm{m}(\bar{\nu}_\mu) = \mathrm{m}(\nu_\mu)
\quad \mathrm{and} \quad \mathrm{m}(\bar{\nu}_e) = \mathrm{m}(\nu_e)\,,
\end{equation}
as in Eq.(9). Adding Eq.(25) and Eq.(24), we find, not considering a possible 
binding energy of the electron to the neutrino lattice of the muon, the energy in the \emph{charged} muon  
\begin{equation}\mathrm{m}(\mu^\pm)\mathrm{c}^2(theor) = 
[3/4\cdot\mathrm{N}\,\mathrm{m}(\nu_\mu)  
+ 3/2\cdot\mathrm{N}\,\mathrm{m}(\nu_e)]\mathrm{c}^2 =
108.395\,\mathrm{MeV}\,, \end{equation}
whereas \hspace{2cm} m($\mu^\pm$)c$^2$(\emph{exp}) = 105.6584\,MeV.

\bigskip
 
  We will now see where the 
discrepancy between Eq.(27) and the experimental value of m($\mu^\pm$)c$^2$ 
comes from. For this we need the energy in the rest mass of e$^\pm$ which is, 
in our model of the electron, according to Eq.(86) in [1], 
\begin{equation}\mathrm{m(e}^\pm)\mathrm{c}^2 = 
\mathrm{N/2\cdot m}(\nu_e)\mathrm{c}^2
= \mathrm{N/2\cdot m}(\bar{\nu}_e)\mathrm{c}^2 \,. \end{equation}
Dividing Eq.(27) by Eq.(28) gives the ratio of the mass of the charged 
muon to the mass of the electron or positron
\begin{equation}\frac{\mathrm{m}(\mu^\pm)}{\mathrm{m(e}^\pm)}(theor)
 = \frac{3}{2}\cdot\frac{\mathrm{m}(\nu_\mu)}{\mathrm{m}(\nu_e)} +3
= \frac{3}{2\alpha_f} + 3 = 208.554 = 1.0086\,\frac{\mathrm{m}(\mu^\pm)}
{\mathrm{m(e}^\pm)}(exp) \,,\end{equation}
using m($\nu_e$) = $\alpha_f$\,$\cdot$\,m($\nu_\mu$), Eq.(4). The term 
3/2$\alpha_f$ in Eq.(29) comes from the
\emph{single} muon or antimuon neutrinos in the cell-sides of the neutrino 
lattice of $\mu^\pm$. In the lattice are then 
N/4\,\,$\cdot$\,$\nu_\mu$($\bar{\nu}_\mu$) neutrinos. With the oscillation 
energy which is, as in the pion, proportional to 
N/2\,$\cdot$\,m($\nu_\mu$)($\bar{\nu}_\mu$), 
the energy in the lattice is proportional to 
3/4\,$\cdot$\,N\,$\cdot$\,m($\nu_\mu$)($\bar{\nu}_\mu$), neglecting m($\nu_e$).
 Divided by m(e$^\pm$) = N/2\,$\cdot$\,m($\nu_e$)($\bar{\nu}_e$) gives, 
with Eq.(4), the 3/2$\alpha_f$ term. The 3/2$\alpha_f$ term in Eq.(29) makes 
the theoretical ratio m($\mu^\pm$)/m(e$^\pm$) independent of the 
values of N, m($\nu_e$) and m($\nu_\mu$). 

   The ratio m($\mu^\pm$)/m(e$^\pm)$ in Eq.(29) differs by +\,2 from 
the empirical formula of Barut [3]
\begin{equation}
 \frac{\mathrm{m}(\mu^\pm)}{\mathrm{m(e}^\pm)}(\emph{emp}) =
\frac{3}{2\alpha_f} +1  = 206.554
 = 0.99896\,\frac{\mathrm{m}(\mu^\pm)}{\mathrm{m(e}^\pm)}(exp)\,,
 \end{equation}
with m($\mu^\pm$)/m(e$^\pm$)(\emph{exp}) = 206.768.
We attribute the difference + 2 between Eq.(29) and Eq.(30)
to the \emph{binding energy of the electron or positron to the neutrino 
lattice of the neutral muon}. That means that
\begin{equation} \Delta\mathrm{E_{\,b}(\mu^\pm_0,e^\pm}) = 
2\,\mathrm{m(e^\pm)c}^2 \,. \end{equation}

\bigskip

   Subtracting from Eq.(27) the binding energy 
$\Delta \mathrm{E\,_b(\mu^\pm_0,e^\pm)}$ = 2\,m(e$^\pm$)c$^2$ = 
N\,m($\nu_e$)c$^2$, we find the total energy in the 
\emph{charged} muons. Simplifying with Eq.(26) we have
\begin{eqnarray} \mathrm{m}(\mu^\pm)\mathrm{c}^2(theor) =
[3/4\cdot\mathrm{Nm}(\nu_\mu) + 1/2\cdot\mathrm{Nm}(\nu_e)]
\mathrm{c}^2 \nonumber\\ 
= 107.35\,\mathrm{MeV} = 1.0160\,\mathrm{m}(\mu^\pm)\mathrm{c}^2(exp) \,,
\end{eqnarray}
whereas m($\mu^\pm)$c$^2$(\emph{exp}) = 105.658\,MeV. We attribute the 
difference between m($\mu^\pm$)c$^2$(theor) and 
m($\mu^\pm$)c$^2$(exp) to the uncertainty of the parameters 
N, m($\nu_\mu$) and m($\nu_e$), all of which are uncertain by about 1\%.

We can, however, eliminate the dependence of Eq.(32) on N, by dividing
Eq.(32) by the mass of e$^\pm$, Eq.(28).
It follows for the ratio of the mass of the 
\emph{charged} muons to the mass of e$^\pm$ that
\begin{equation}\frac{\mathrm{m}(\mu^\pm)}{\mathrm{m(e}^\pm)}(theor)
= \frac{3}{2}\cdot\frac{\mathrm{m}(\nu_\mu)}{\mathrm{m}(\nu_e)} + 1\,,
\end{equation}
which does not depend on N any more, but still on m($\nu_\mu$) and 
m($\nu_e$).
We now eliminate also the uncertainty of the neutrino masses through
m($\nu_e$) = $\alpha_f$\,$\cdot$\,m($\nu_\mu$) (Eq.4), and find that
\begin{equation} \frac{\mathrm{m}(\mu^\pm)}{\mathrm{m(e}^\pm)}(theor) = 
\frac{3}{2\alpha_f} + 1 = 206.554 = 
0.99896\,\frac{\mathrm{m}(\mu^\pm)}{\mathrm{m(e}^\pm)}(exp)\,,
 \end{equation}
which is equal to Barut's formula, Eq.(30). The experimental ratio
\newline
m($\mu^\pm$)/m(e$^\pm$)(\emph{exp}) = 206.768 is 1.00104 times larger than 
our theoretical value for m($\mu^\pm$)/m(e$^\pm$) = 206.554, which we find 
when we take the binding energy into 
account. The fine structure constant $\alpha_f$ in Eq.(34) is an exactly 
known constant, so the theoretical value of m($\mu^\pm$)/m(e$^\pm$)
is very firm. Taking the binding of the electron to the neutrino lattice 
of the muon into account we have explained from our model of the muon the
empirical formula for the ratio m($\mu^\pm$)/m(e$^\pm$) suggested 
by Barut [3], following an earlier suggestion by Nambu [2]. Barut's 
formula is accurate to about one part in a thousand. Eq.(34) means
also that we have determined theoretically the rest mass of the muon with 
an accuracy of one part in a thousand. 

\bigskip    

   We have thus shown that the binding energy of the electric charge 
e$^\pm$ brought into the neutrino lattice of the neutral muons  
is 2\,m(e$^\pm$)c$^2$. We have also found from equations (14) and (31) 
that the binding energy of an  
electron or positron in the $\pi^\pm$ lattice is twice the binding energy
of e$^\pm$ in the $\mu^\pm$ lattice. The difference of both binding 
energies appears to be a consequence of the presence 
of only one muon neutrino or anti-muon neutrino in a side of each 
cell of the muon lattice, whereas one muon neutrino \emph{and} one anti-muon 
neutrino are in the sides of each cell of the pion lattice.

\section*{Conclusions}

   We have been concerned with the consequences of the presence of
the electric charge e$^\pm$ in the mass of the muon $\mu^\pm$ and 
the pion $\pi^\pm$, and with the explanation of the experimental ratios 
of the masses of $\mu^\pm$ and $\pi^\pm$ to the mass of the electron or 
positron,

\bigskip
\noindent
\hspace{2.2cm}m($\mu^\pm$)/m(e$^\pm$) = 206.7683 = 
1.00104\,$\cdot$\,(3/2$\alpha_f$ + 1),\\
\smallskip
\hspace{2.2cm}m($\pi^\pm$)/m(e$^\pm$) = 273.1320 =
1.00022\,$\cdot$\,(2/$\alpha_f$ $-$ 1)\,.

\bigskip

In both cases the terms on the right hand side of the equations with half 
integer or integer multiples of 1/$\alpha_f$ agree exactly with what one 
expects from our model of the particles consisting of lattices  
of muon neutrinos, anti-muon neutrinos, electron neutrinos and anti-electron
neutrinos, whose masses are related by m($\nu_e$) = 
$\alpha_f$\,$\cdot$\,m($\nu_\mu$). 

   From the ratios of the measured particle masses to the mass of the 
electron or positron follows furthermore that integer multiples of 
1/$\alpha_f$ are in

\bigskip

\noindent
\hspace{2.35cm}m(K$^\pm$)/m(e$^\pm$) = 966.102 =
0.99984\,$\cdot$\,(7/$\alpha_f$ + 7)\,,\\
\smallskip
\hspace{2.35cm}m(K$^0$)/m(e$^\pm$) = 973.873 =
0.99961\,$\cdot$\,(7/$\alpha_f$ + 15)\,,\\
\smallskip
\hspace{2.2cm} m(n)/m(e$^\pm$) = 1838.684 =
1.000098\,$\cdot$\,(14/$\alpha_f$ $-$ 80)\,,\\ 
\smallskip
\hspace{2.2cm} m(p)/m(e$^\pm$) = 1836.153 = 
1.000081\,$\cdot$\,(14/$\alpha_f$ $-$ 82.5)\,,\\
\smallskip
\hspace{2.2cm} m(D$^\pm$)/m(e$^\pm$) = 3658.13 = 
0.997\,$\cdot$\,(28/$\alpha_f$ $-$ 168)\,,

\bigskip
\noindent
as it should be if m(n) $\cong$ 2\,m(K$^0$), and
as it should be if m(D$^\pm$) $\cong$ m(p + $\bar{\mathrm{n}}$).

   The integer numbers added to the 1/$\alpha_f$ 
terms in the mass ratios above do, however, not agree, in each of the 
mass ratios listed, with what one expects from our model of
the neutral neutrino bodies of the particles. These differences are caused 
by the introduction of the electric charges e$^\pm$ into 
the neutrino bodies of the particles. The introduction of e$^\pm$ into 
the neutral body of a particle changes the number of the electron neutrinos
or anti-electron neutrinos in the particles, not the number of the muon 
neutrinos or anti-muon neutrinos.

   When an electric charge e$^\pm$  is introduced into 
the large neutrino bodies of the pion and muon 
an exothermic binding energy is released, which amounts in the case
of $\mu^\pm$ to 2\,m(e$^\pm$)c$^2$, and in the case of $\pi^\pm$ 
to 4\,m(e$^\pm$)c$^2$. 
We have learned that the ratio of the mass of the charged muons to the  
mass of the electron or positron can be explained with an accuracy of 
about 1\,$\cdot$\,10$^{-3}$, and that the ratio of the mass of the charged 
pions to the mass of the electron or positron can be explained  with an 
accuracy of 2.2\,$\cdot$\,10$^{-4}$, if the binding energy of the charge to
the neutrino lattices of the muons and pions is taken into account.
  
   We have also learned that the electron neutrinos or anti-electron neutrinos,
as well as the charge elements, brought by the charge into the simple cubic 
neutrino lattice of the pions must sit in the center of the sides of the cubic cells of the neutrino lattice of the pions, making them face-centered cubic cells, which is essential for the stability of the $\pi^\pm$\,mesons. It seems 
likely that the electron or anti-electron neutrinos brought by e$^\pm$ 
into the muon lattice occupy the places vacated by the emission  
of the muon neutrinos or anti-muon neutrinos in the $\pi^\pm$ 
decay, that means that the muon lattice is cubic. The charge elements brought also with e$^\pm$ into the neutral lattice of the muons make the cells in the charged muons face-centered cubic.

\section*{References}

[1] Koschmieder,\,E.L. http://arXiv:0804.4848v8 (2013).

\smallskip
\noindent
[2] Nambu,\,Y. Prog.Th.Phys. {\bfseries7},595 (1952).

\smallskip
\noindent
[3] Barut,\,A.O. Phys.Lett.B. {\bfseries73},310 (1978).

\smallskip
\noindent
[4] Born,\,M. Proc.Camb.Phil.Soc. {\bfseries36},160 (1940).

\vspace{1.0cm}

\end{document}